\documentclass[%
 reprint,
superscriptaddress,
 amsmath,amssymb,
 aps,prl,
]{revtex4-1}

\usepackage{amsmath}
\usepackage{mathtools}
\usepackage{physics}
\usepackage{bbold}
\usepackage{tikz}
\usepackage{graphicx}
\usepackage{dcolumn}
\usepackage{bm}
\usepackage{placeins}

\usepackage[]{hyperref}

\newcommand{\nvec}[1]{\boldsymbol{#1}}

\def \FigureOne
{
\begin{figure}[!t]
\includegraphics[width=\linewidth]
{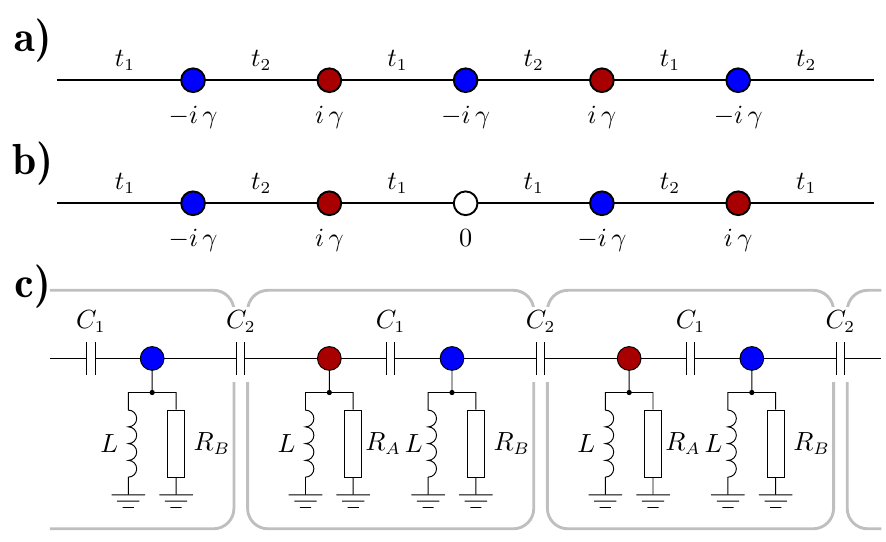}
 \caption{Theoretical hopping model and circuit diagram. a) Bulk model with two sites per unit cell (red and blue), with alternating hoppings $t_1$ and $t_2$ and on-site gain/loss terms $\pm \mathrm{i} \gamma$. b) Insertion of the $\mathcal{PT}$ symmetric defect. c) Circuit diagram of the experimental implementation, unit cells framed in grey boxes. The hoppings are realized by capacitors $C_1$, $C_2$, the on-site gain and loss by resistive elements $R_A$, $R_B$. The inductor $L$ tunes the resonance frequency of the circuit.
 }
 \label{fig:circuit}
\end{figure}
}

\def\FigureTwo
{
\begin{figure*}[!t]
    \centering
    \includegraphics[width=\linewidth]
    {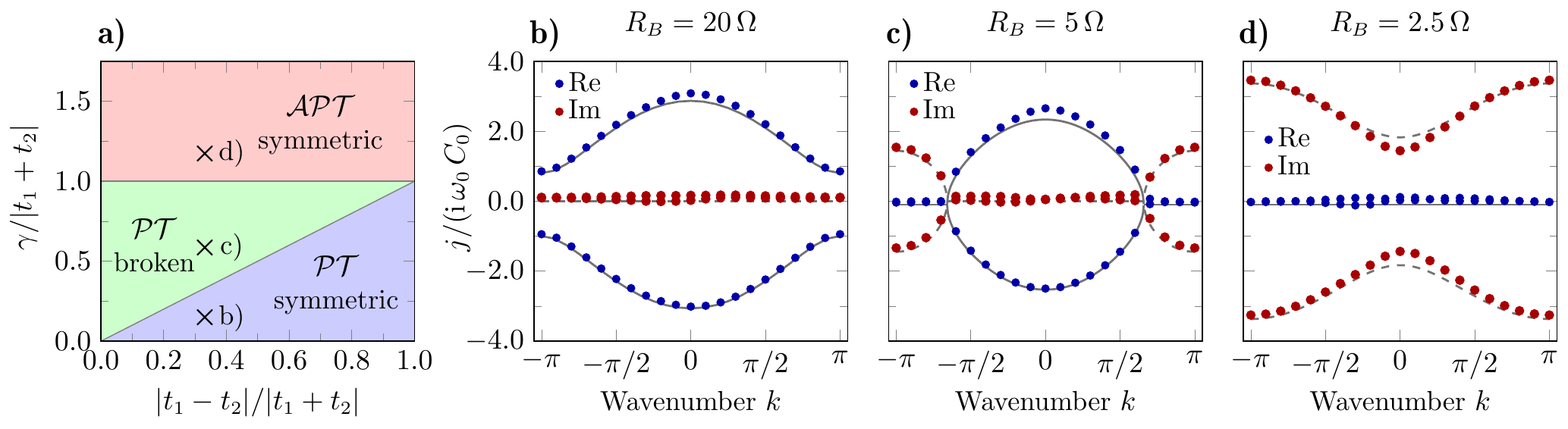}
    \caption{Admittance band structure of the periodic $\mathcal{PT}$ SSH chain (20 unit cells) in different symmetry phases. a) Phase diagram of the $\mathcal{PT}$ SSH model. b) -- d) Theoretical and measured admittance band structure for $C_1 = 100\,\mathrm{nF}$, $C_2 = 200\,\mathrm{nF}$, $L = 10 \, \mathrm{\mu H}$, $(R_A)^{-1}=0$ 
    and three different values of $R_B$, measured at $\omega_0=90.6\,\mathrm{kHz}$. 
    The blue points (solid lines) show the real part of the measured (theoretical) normalized admittance eigenvalues $j(k)/(\mathrm{i}\omega_0 C_0)$ with $C_0=100\,\mathrm{nF}$, the red points (dashed lines) the imaginary parts. The values are corrected by the nominal loss offset of $(2\,R_B)^{-1}$.
    }
\label{fig:bands}
\end{figure*}
}

\def\FigureThree
{
\begin{figure*}[!t]
    \centering
    \includegraphics[width=\linewidth]{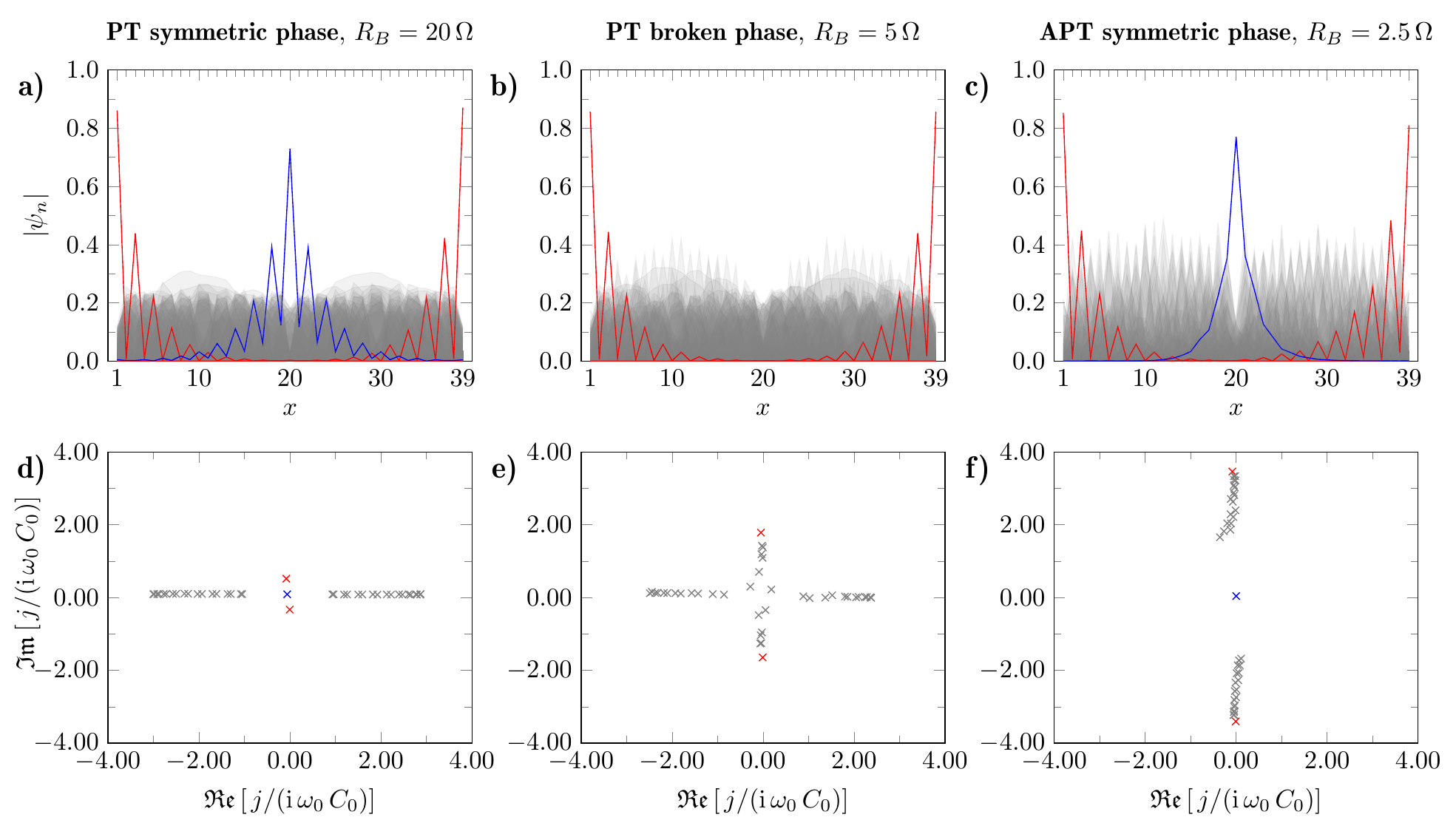}
    \caption{Measured eigenvectors and eigenvalues of the $\mathcal{PT}$ SSH chain (39 sites) with defect site and open boundary conditions. The edge states are marked in red, the defect state in blue and the bulk states in grey. $C_1$, $C_2$, $L$ and $R_A$ are identical to the PBC measurement in \hyperref[fig:bands]{Fig.2}.
    a) -- c) Absolute values of the measured admittance eigenvectors. d) -- f) Measured admittance eigenvalues in the complex plane. The values are corrected by the nominal loss envelope of $(2\,R_B)^{-1}$.}
    \label{fig:OBC}
\end{figure*}
}

\begin{document}


\title{Topological defect engineering and $\mathcal{PT}$-symmetry in non-Hermitian electrical circuits}

\author{Alexander Stegmaier}
 \affiliation{Institute for Theoretical Physics and Astrophysics, University of W\"urzburg, Am Hubland, D-97074 W\"urzburg, Germany}
 \author{Stefan Imhof}
 \affiliation{Physikalisches Institut and R\"ontgen Research Center for Complex Material Systems, Universit\"at W\"urzburg, D-97074 W\"urzburg, Germany}
\author{Tobias Helbig}
 \affiliation{Institute for Theoretical Physics and Astrophysics, University of W\"urzburg, Am Hubland, D-97074 W\"urzburg, Germany}
 \author{Tobias Hofmann}
 \affiliation{Institute for Theoretical Physics and Astrophysics, University of W\"urzburg, Am Hubland, D-97074 W\"urzburg, Germany}
 \author{Ching Hua Lee}
 \affiliation{Department of Physics, National University of Singapore, Singapore 117542}
 \author{Mark Kremer}
 \affiliation{Institut f\"ur Physik, Universität Rostock, Albert-Einstein-Straße 23, 18059 Rostock}
 \author{Alexander Fritzsche}
 \affiliation{Institut f\"ur Physik, Universität Rostock, Albert-Einstein-Straße 23, 18059 Rostock}
 \affiliation{Institute for Theoretical Physics and Astrophysics, University of W\"urzburg, Am Hubland, D-97074 W\"urzburg, Germany}
 \author{Thorsten Feichtner}
 \affiliation{Physikalisches Institut and R\"ontgen Research Center for Complex Material Systems, Universit\"at W\"urzburg, D-97074 W\"urzburg, Germany}
 \author{Sebastian Klembt}
 \affiliation{Technische Physik and Wilhelm-Conrad-Röntgen-Research Center for Complex Material Systems, Universität Würzburg, Am Hubland, D-97074 Würzburg, Germany}
 \author{Sven H\"ofling}
 \affiliation{Physikalisches Institut and R\"ontgen Research Center for Complex Material Systems, Universit\"at W\"urzburg, D-97074 W\"urzburg, Germany}
 \author{Igor Boettcher}
 \affiliation{Joint Quantum Institute, University of Maryland, College Park, MD 20742, USA}
 \author{Ion Cosma Fulga}
 \affiliation{Institute for Theoretical Solid State Physics, IFW Dresden, 01171 Dresden, Germany}
 \author{Oliver G. Schmidt}
 \affiliation{Institute for Integrative Nanosciences, Leibniz IFW Dresden, 01069 Dresden, Germany}
 \author{Martin Greiter}
 \affiliation{Institute for Theoretical Physics and Astrophysics, University of W\"urzburg, Am Hubland, D-97074 W\"urzburg, Germany}
 \author{Tobias Kiessling}
 \affiliation{Physikalisches Institut and R\"ontgen Research Center for Complex Material Systems, Universit\"at W\"urzburg, D-97074 W\"urzburg, Germany}
 \author{Alexander Szameit}
 \affiliation{Institut f\"ur Physik, Universität Rostock, Albert-Einstein-Straße 23, 18059 Rostock}
 \author{Ronny Thomale}
 \affiliation{Institute for Theoretical Physics and Astrophysics, University of W\"urzburg, Am Hubland, D-97074 W\"urzburg, Germany}

\date{\today}

\begin{abstract}
We employ electric circuit networks to study topological states of matter in non-Hermitian systems enriched by parity-time symmetry $\mathcal{PT}$ and chiral symmetry anti-$\mathcal{PT}$ ($\mathcal{APT}$). The topological structure manifests itself in the complex admittance bands which yields excellent measurability and signal to noise ratio. We analyze the impact of $\mathcal{PT}$ symmetric gain and loss on localized edge and defect states in a non--Hermitian Su--Schrieffer--Heeger (SSH) circuit. We realize all three symmetry phases of the system, including the $\mathcal{APT}$ symmetric regime that occurs at large gain and loss. We measure the admittance spectrum and eigenstates for arbitrary boundary conditions, which allows us to resolve not only topological edge states, but also a novel $\mathcal{PT}$ symmetric $\mathbb{Z}_2$ invariant of the bulk. We discover the distinct properties of topological edge states and defect states in the phase diagram. In the regime that is not $\mathcal{PT}$ symmetric, the topological defect state disappears and only reemerges when $\mathcal{APT}$ symmetry is reached, while the topological edge states always prevail and only experience a shift in eigenvalue. Our findings unveil a future route for topological defect engineering and tuning in non-Hermitian systems of arbitrary dimension.\\
\end{abstract}

\pacs{Valid PACS appear here}
\maketitle


\paragraph*{Introduction.}
Topological matter has become a prime subject of contemporary research across a wide range of areas in physics~\cite{Hasan_2010, Qi_2011, Lu_2014}. As they feature a certain degree of robustness and insensitivity against perturbations, topological boundary states promise various future
technological applications. These include unidirectional transport without backscattering \cite{Wang_2009}, wave guidance, robust quantum computation \cite{Nigg_2014,Blanco-Redondo568}, or other functionalities that, due to their topological character, are tolerant to fabrication imperfections and unwanted parasitic effects. Furthermore, topological properties experience an intriguing degree of diversification when they are combined with discrete symmetries such as parity $\mathcal{P}$ and time reversal $\mathcal{T}$. Combined, $\mathcal{PT}$ symmetry descended from the mathematical analysis of quantum field theories and triggered the first wave of interest in non-Hermitian photonics even before the perspective of topological matter in synthetic platforms had been shaped \cite{El-Ganainy:07,PhysRevLett.101.080402,Makris_2008,Longhi_2009}. $\mathcal{PT}$ symmetry enables the emergence of a plethora of phenomena such as non-reciprocal light propagation \cite{Feng_2011, Weidemann_2020}, 
unidirectional invisibility \cite{Regensburger_2012} and arbitrarily fast state evolution despite limited bandwidth \cite{Ramezani_2012} not only in one but also higher dimension~\cite{kremerli}. The presence of $\mathcal{P}$ and $\mathcal{T}$ symmetry ensures a real spectrum \cite{Bender_1998, Bender_1999, Mostafazadeh_2002} and hence conserved dynamics despite the presence of gain and loss. 

Topolectric circuits \cite{Ningyuan_2015, Albert_2015, Lee_2018} have only recently emerged as a platform for synthetic topological matter. Building on the insight that Berry's phase only requires a physical system to exhibit the principle of interference, topological band structures can result from systems as diverse as mechanics described by Newton's equation \cite{Nash_2015, Huber_2016}, optical waveguides by the Helmholtz equation \cite{Lu_2014}, and electrons by the Schr\"odinger equation~\cite{Berry_1984}. For topolectric circuits, it is the combination of Kirchhoff's rule and the complex-valued admittance which provides the source for non-trivial topological circuit networks~\cite{Lee_2018}. As opposed to other topological platforms where topological properties predominantly emerge through energy or frequency bands, a topolectric circuit network fed by an external
AC current exhibits topological admittance bands. Because of the unparalleled precision and simplicity of the required voltage measurements of the circuit nodes
against ground, electric circuits enable a detailed design and measurement of admittance bands \cite{Helbig_2019}. Due to the large variety of circuit components, the range of model Hamiltonians realized in topolectric systems includes Chern insulators \cite{Hofmann_2019,PhysRevB.99.235110,PhysRevB.100.081401,yang2020observation}, non-reciprocal chains and other non-Hermitian phenomena \cite{Yao_2018,PhysRevB.99.201103,Helbig_2020,PhysRevB.100.054301,PhysRevResearch.2.023265,PhysRevLett.124.046401,PhysRevResearch.2.022062, li2020impurity}, higher-order topological insulators \cite{Imhof_2018,bahl,PhysRevB.98.201402,PhysRevB.100.201406}, topological semimetals \cite{Lee_2018,arsch,PhysRevB.99.020302,lee-nc2020,Rafi_2020}, entanglement simulators for photons~\cite{gorlach}, and topological states in arbitrary dimension larger than three \cite{Lee_2018,yidong}.

In this Letter, we accomplish two things. First, we design and implement a circuit realization of the non-Hermitian Su-Schrieffer-Heeger (SSH) chain where gain and loss can be tuned to investigate the $\mathcal{PT}$ symmetry. Through measuring the non-Hermitian SSH circuit admittance spectrum and eigenstates for periodic and open boundary conditions, we can experimentally access both the topological midgap states and the topological bulk invariant. Second, we are able to analyze the distinct nature of the topological SSH edge and defect states upon tuning gain and loss. While the topological edge states are only modified in terms of their decay length, we observe the disappearance and reentrance of the topological defect state depending on the symmetry of the non-Hermitian circuit.

\FigureOne

\FigureTwo

\paragraph*{$\mathcal{PT}$ SSH circuit.} The $\mathcal{PT}$ SSH tight-binding model, schematically illustrated in \mbox{\hyperref[fig:circuit]{Fig. 1a}}, consists of a chain with alternating hoppings $t_1$, $t_2$, and an alternating on-site gain and loss term $\pm \, \mathrm{i} \, \gamma$.
The Hermitian SSH model for $\gamma=0$, which was initially reported to be realized in an optical experiment in Ref.~\onlinecite{malkova}, is symmetric under both parity  $\mathcal{P}$, which acts as inversion along a bond, and time reversal $\mathcal{T}$, which acts as complex conjugation. The addition of gain/loss breaks both $\mathcal{P}$ and $\mathcal{T}$, as they change the sign of $\gamma$. When combined, $\mathcal{PT}$ is still conserved in the non-Hermitian SSH model.
Chiral symmetry, represented by $\sigma_z$ in the SSH band basis in Eq.(\ref{eq:Lap}), is likewise violated by the gain/loss term. Combining it with time reversal $\mathcal{T}$, however, yields a symmetry in the non-Hermitian case, the anti-$\mathcal{PT}$ symmetry \mbox{${\mathcal{APT} =\sigma_z \mathcal{T}}$ \cite{Longhi_2018, Li_2013, Yang_2017}.}
The SSH bulk is characterized by a $\pi$-quantized Berry phase, the Zak phase \cite{Zak_1989}, that predicts the presence of mid-gap edge states by Hermitian bulk-boundary correspondence (BBC).
For non-zero gain and loss, the edge states individually always break $\mathcal{PT}$ symmetry since $\mathcal{PT}$ maps them to the opposite edge \cite{Esaki_2011, Hu_2011}. Even though the bulk-boundary correspondence is challenged in its generality, topological invariants can often be transcribed from the Hermitian to the non-Hermitian case.

In order to design a manifestly $\mathcal{PT}$ symmetric topological midgap state in a waveguide system, Poli et al. \cite{poli} and Weimann et al. \cite{Weimann_2016} suggested to insert a defect site that forms a kernel of $\mathcal{PT}$ symmetry, as shown in \mbox{\hyperref[fig:circuit]{Fig. 1b}}. This way, a domain wall between different topological phases is created where a $\mathcal{PT}$ symmetric topological midgap state can emerge.
Weimann et al. observed that the localized defect state disappears when the bulk states break $\mathcal{PT}$ symmetry. As we will demonstrate, there is another symmetry phase transition, where all eigenstates of the non-Hermitian SSH model become $\mathcal{APT}$ symmetric. Here, the defect state exhibits reentrant behavior. Reaching this regime requires a large gain/loss term which did not prove feasible in a platform such as waveguides (Appendix C).

We adopt electric circuits to experimentally observe
all three regimes, i.e., the $\mathcal{PT}$, $\mathcal{APT}$, and the symmetry-broken phase, and investigate the evolution of the topological defect and edge states. 
The non-Hermitian SSH model is represented by the admittance matrix, also termed circuit Laplacian $J(\omega)$ \cite{Lee_2018} of the circuit. 
The Laplacian describes the voltage response $\nvec{V}(\omega)$ to an AC input current $\nvec{I}(\omega)$ according to
\begin{align}
  \nvec{I}(\omega) = J(\omega) \, \nvec{V}(\omega). \label{eq:Lap}
\end{align}
The vector components of $\nvec{V}$ and $\nvec{I}$ correspond to the nodes or sites in the circuit. 
We investigate the circuit in a steady state, such that the AC frequency $f = \omega/(2\pi)$ can be treated as a parameter.

Equation (\ref{eq:Lap}) implies that admittance measurements in dissipative circuits are as precise as in those without loss, since the admittance of resistive and reactive circuit components only differs by a complex phase. Here, circuits supersede platforms that rely on transport measurements, where loss causes exponentially diminished signal intensities worsening the obtainable signal-to-noise ratio. 
\mbox{\hyperref[fig:circuit]{Fig. 1c}} shows our SSH circuit design. The hoppings are represented by capacitors
between neighboring nodes, the onsite gain and loss by resistive elements to ground. While not further discussed in the main text, balanced gain and loss can be implemented using negative impedance converters, see \mbox{Appendix G}. 
Connecting all nodes to ground through inductors allows us to tune the offset of the circuit's admittance by adjusting the AC frequency $\omega$. 
With periodic boundary conditions, the Laplacian can be expressed in Bloch form for the two-site unit cell \mbox{(Appendix A, B)} as
\begin{align}
 J(k, \omega) = \mathrm{i} \,\omega \, C_0 \big[(-t_1-t_2\,\cos{k})\sigma_x - t_2\,\sin{k}\,\sigma_y + \nonumber \\
 + \mathrm{i} \,\gamma(\omega) \, \sigma_z + \mathrm{i} \, \epsilon(\omega) \, \mathbb{1} \big].
\end{align}
Here, $\sigma_i$ denotes the $i$th $(2\times2)$ Pauli matrix, $t_1$ ($t_2$) is the intra-cell (inter-cell) hopping, and $\gamma$ represents the alternating gain/loss. If the input current is tuned to the mid-gap frequency ${\omega_0=[(t_1 + t_2)\,C_0\, L]^{-1/2}}$, $\epsilon$ merely describes a constant resistive offset (Appendix B), which we will omit in the following.
In what follows, we use $t_1=1$, $t_2=2$ and always set $\omega = \omega_0$ unless stated otherwise.

\FigureThree

\paragraph*{Topological edge modes and non-Hermitian bulk-boundary correspondence.}
We measure the admittance band structure of the SSH circuit in all symmetry regimes. 
At first, the circuit is set up with periodic boundaries and no defect site. 
With a loss envelope of $\epsilon = -\gamma$, the resistive term vanishes on one site of the unit cell, such that only every other node is connected to ground by a resistor $R_B = 1 / ( 2 \omega_0 \, C_0 \, \gamma)$.
The phase diagram of the $\mathcal{PT}$ SSH model shown in \mbox{\hyperref[fig:bands]{Fig. 2a}} depends on the dimerization $\abs{t_1-t_2}$ and the gain/loss term $\gamma$.
The phase transitions can be understood by investigating the admittance band structure
\begin{align}
 j(k)/(\mathrm{i}\, \omega_0 \, C_0) = \pm \sqrt{t_1^2 + t_2^2 + 2 t_1 t_2 \cos{k} - \gamma^2} \label{eq:bands}
\end{align}
that is formed by the eigenvalues of the Laplacian (\ref{eq:Lap}). For sufficiently small $\gamma$, the radicant is always non-negative, so that the resulting bands are real-valued, as depicted in \mbox{\hyperref[fig:bands]{\mbox{Fig. 2b}}} for $R_B=20\, \Omega, \gamma = 0.433$. As discussed in detail in Appendix D, this means that all eigenstates are $\mathcal{PT}$ symmetric, hence defining the $\mathcal{PT}$ symmetric phase. For large values of $\gamma$, the radicant is always negative, so that the bands are fully imaginary, which is represented by the measurement at $R_B=2.5\, \Omega, \gamma=3.46$ in \mbox{\hyperref[fig:bands]{Fig. 2d}}. Accordingly, the eigenstates are $\mathcal{APT}$ symmetric.
The $\mathcal{PT}$ broken phase occurs for intermediate $\gamma$, where the radicant of (\ref{eq:bands}) is only negative for a part of the Brillouin zone. As seen in \mbox{\hyperref[fig:bands]{Fig. 2c}}, at $R_B=5\, \Omega, \gamma=1.73$, exceptional points \cite{Berry_2004, Miri_2019} appear in the band structure where the radicant of (\ref{eq:bands}) changes sign and the real and imaginary part of both bands is zero.

Since our measurements allow us to reconstruct the eigenstates of $J(k)$, we can calculate topological bulk winding numbers of our circuit. 
The Zak phase of the $\mathcal{PT}$ SSH model is only quantized in the $\mathcal{PT}$ symmetric regime \cite{Liang_2013}. Calculating it from the {\it measured} eigenstates of the lower band, we obtain $(-0.999 \pm 0.020)\,\pi$ for the $\mathcal{PT}$ symmetric setup,
$(0.651\pm0.021)\,\pi$ for the $\mathcal{PT}$ broken setup,
and $(0.208\pm0.029)\,\pi$ for the $\mathcal{APT}$ symmetric setup. 
These result match the theoretical  values of $\pm \pi$, $0.655 \, \pi$ and $0.197\,\pi$. It confirms that the $\pi$-quantization is present in the $\mathcal{PT}$ symmetric case but absent for the other two.

For this reason, we propose a generalized classification of the non-Hermitian SSH bulk invariant based on $\mathcal{PT}$ symmetry, the $\mathcal{PT}$ winding number 
\begin{align}
 \mathcal{W}_n^\mathcal{PT} = \frac{1}{2 \pi} \oint_{\mathrm{BZ}} \!\! \mathrm{d}k\, \partial_k \arg{\left\{\Psi^\dagger_n(k)\,\mathcal{PT}\,\Psi_n(k)\right\}}.
 \end{align}
Here, $\Psi_n(k)$ denote the eigenstates of the $n$th band. If these states are $\mathcal{PT}$ symmetric, they acquire a complex phase under application of  the $\mathcal{PT}$ operator, $\mathcal{PT}\,\Psi_n(k)=\mathrm{e}^{\mathrm{i}\varphi_n(k)} \Psi_n(k)$. In this case, the winding number (4) corresponds to the winding of $\mathrm{e}^{\mathrm{i}\varphi_n(k)}$ around the origin of the complex plane. We show in Appendix D, that this winding number times $\pi$ is equal to the Zak phase. This equivalence breaks down in the $\mathcal{PT}$ broken and $\mathcal{APT}$ symmetric phase. 
Whereas the Zak phase loses its quantization outside the $\mathcal{PT}$ symmetric domain, $\mathcal{W}^\mathcal{PT}$ still provides a $\mathbb{Z}_2$ classification and predicts the presence of edge states via BBC.
For our three measured setups, the $\mathcal{PT}$ winding number always evaluates to one, predicting the presence of edge states in all three phases. This is confirmed in our measurements of an open boundary setup, see \mbox{\hyperref[fig:OBC]{Fig. 3a-c}}.
For the $\mathcal{PT}$ SSH model, the classification by the $\mathcal{PT}$ winding number gives the same result as the Q-matrix approach in non-Hermitian systems \cite{Esaki_2011, Yao_2018}.

\paragraph*{Topological defect mode.}
In order to study the influence of dissipative terms onto topological midgap states, we apply open boundary conditions and insert the $\mathcal{PT}$ symmetric defect site into the circuit. We study the same points of parameter space as for PBC.
The combined action of $\mathcal{PT}$ and $\mathcal{APT}$ yields a linear anti-commuting symmetry of the system. This constrains the eigenvalues to appear in pairs of opposite sign \mbox{(Appendix D)}. Upon addition of the defect site, the circuit has an uneven number of sites, and as such a state with eigenvalue zero should exist. 
In our measurements, we identify a defect midgap state in the $\mathcal{PT}$ and $\mathcal{APT}$ symmetric regimes for $R_B=20\,\Omega$ and $R_B=2.5\,\Omega$, respectively, marked in blue in \mbox{\hyperref[fig:OBC]{Figs. 3a, 3c}} \footnote{A code which numerically reproduces the spectra, eigenstates, and bandstructures of the $\mathcal{PT}$ SSH chain is available at \href{https://doi.org/10.5281/zenodo.4239893}{https://doi.org/10.5281/zenodo.4239893}}. We confirm that it is located at zero admittance, see \mbox{\hyperref[fig:OBC]{Figs. 3d,f}}. In the $\mathcal{PT}$ broken phase \mbox{(\hyperref[fig:OBC]{Figs. 3b,e})} at $R_B=5\,\Omega$, no localized defect state is present. 
The selective appearance of the defect midgap state is contrasted by the edge midgap states which we observe in all three symmetry phases \mbox{(\hyperref[fig:OBC]{Fig. 3}, marked red)}. As seen in \mbox{\hyperref[fig:OBC]{Fig. 3d-f}}, their eigenvalues are fully imaginary, so they are symmetric under $\mathcal{APT}$.

Exponentially localized states in a periodic lattice can be characterized by a complex-valued wave number $k\in \mathbb{C}$. Their eigenvalue is then determined by the corresponding analytic continuation of the band structure (Appendix F).
Reversing this relationship, we find the localization length of the mid-gap state by solving $\det{J(k)}\stackrel{!}{=}0$ and obtain 
\begin{align}
k^{\pm} = \pm \arccos{\left(\frac{\gamma^2-t_1^2-t_2^2}{2 t_1 t_2}\right)}.\label{eq:EPs}
\end{align}
Herein, $k^\pm$ are the two exceptional points of $J(k)$ in complex momentum space, i.e. the branch points of \mbox{equation (\ref{eq:bands}).}
The localization length of the defect state is given by $\abs{\mathfrak{Im}\{ k^\pm\}}$. Since the arccos is real-valued for arguments between $-1$ and $1$, but complex otherwise, we expect that the defect state is delocalized in the former case but localized in the latter. 
In the $\mathcal{PT}$-symmetric phase, $\abs{\gamma} < \abs{t_1-t_2}$ and $k^\pm$ take the form ${\pi \pm \mathrm{i} \kappa}, \; \kappa \in \mathbb{R}^+$. This implies that the defect state is localized with inverse localization length $\kappa$, as seen in \mbox{\hyperref[fig:OBC]{Fig. 3a}}.
In the $\mathcal{PT}$ broken phase, the $k^\pm$ are real-valued. Therefore, they appear in the band structure as the two exceptional points shown in \mbox{\hyperref[fig:bands]{Fig. 2c}}. Since these modes are delocalized, no defect state can be observed, see \mbox{\hyperref[fig:OBC]{Fig. 3b, e}}.
In the $\mathcal{APT}$ symmetric phase, $\abs{\gamma} > \abs{t_1 + t_2}$, $k^\pm$ is fully imaginary, $k^\pm = \pm \mathrm{i} \kappa$. Consequently, the defect state exhibits reentrant behavior in this regime, as seen in \mbox{\hyperref[fig:OBC]{Fig. 3c, f}}.

\paragraph*{Conclusion.}
By the prototypical example of a $\mathcal{PT}$ SSH circuit for which we can access all symmetry regimes of the phase diagram, we propagate electric circuits as a preeminently suited platform to study the interdependence of symmetry, non-Hermiticity, and topology. In particular, we find that topological defect engineering might enable aspects of topological mode tuning inaccessible to topological edge modes. Topolectric circuits will prove promising to enter this unchartered territory of topological matter with an unprecedented degree of tunability and measurability. For instance, due to the circuit platform, the principles laid out here can be readily generalized to higher dimensional circuits, and can be combined with non-linearities~\cite{alualu,kotwal2019active,doi:10.1063/1.5142397}.

\paragraph*{Acknowledgments.}  We thank K. Leo and M. Segev for discussions. This work is funded by the Deutsche Forschungsgemeinschaft (DFG, German Research Foundation) through Project-ID 258499086 - SFB 1170 and through the W\"urzburg-Dresden Cluster of Excellence on Complexity and Topology in Quantum Matter-ct.qmat Project-ID 390858490 - EXC 2147. T.He. was supported by a Ph.D. scholarship of the Studienstiftung des deutschen Volkes. IB is funded by the DoE BES Materials and
Chemical Sciences Research for Quantum Information Science program (award No. DE-SC0019449), NSF PFCQC program, DoE ASCR (award No. DE- SC0020312 and DE-SC0019040), AFOSR, ARO MURI, ARL CDQI, AFOSR MURI, and NSF PFC at JQI.


%

\end{document}